\documentclass[aps,12pt,pra,final,eqsecnum]{revtex4}
\usepackage{amsmath,amssymb,latexsym}
\usepackage{epsfig,rotating,graphicx}

\begin{document}
\begin{center}
{\large \textbf{Calculating state-to-state transition probabilities within TDDFT}} \vspace{0.2in}
 
Nina~Rohringer,Simone~Peter,Joachim~Burgd\"orfer
 
Institute for Theoretical Physics, Vienna University of Technology, A-1040
Vienna, Austria\\[2ex]
 
\vspace{0.25in}
\textbf{Abstract}\\
The determination of the elements of the S-matrix within the framework of 
time-dependent density-functional theory (TDDFT) has remained a widely open 
question. We explore two different methods to calculate state-to-state transition 
probabilities. The first method closely follows the extraction of the S-matrix from 
the time-dependent Hartree-Fock approximation. This method suffers from 
cross-channel correlations resulting in oscillating transition probabilities in 
the asymptotic channels. An alternative method is proposed which corresponds to 
an implicit functional in the time-dependent density. It gives rise to stable and 
accurate transition probabilities. An exactly solvable two-electron system serves 
as benchmark for a quantitative test. 
\end{center}
                                                                                                                   
\date{\today}
As a matter of principle, time-dependent density functional theory \cite{rungegross}
provides a highly efficient method to solve the time-dependent quantum many-body problem. 
It yields directly the time-dependent one-particle density $n(\vec{r},t)$ 
of the many-body system. All physical observables of the quantum system can, in principle, 
be determined from the density. In practice there a two essential ingredients to a TDDFT
calculation. First an approximation to the time-dependent exchange-correlation potential 
$V_{xc}[n](\vec{r},t)$ has to be found which via the non-interacting Kohn-Sham system 
determines the evolution of the density. The second ingredient are functionals 
that allow the extraction of physical observables from the density.
For some of the observables such as the ground-state energy extraction is straight forward 
within ground-state density functional theory \cite{hohenbergkohn}.
Excited-state spectra have been obtained from linear-response functionals \cite{casida,lr}. Beyond
linear response, the time-dependent dipole moment which governs the emission of high-harmonic
radiation can be directly determined from $n(\vec{r},t)$. Ionization probabilities can be approximately 
extracted by identifying the integrated density beyond a certain critical distance from the bound system
with the flux of ionized particles \cite{lappas,petersilka}. However, in general,  on the most fundamental level,
state-to-state transition probabilities contain the full information on the response of a many-body
system to an external perturbation. One example are bound-bound transition amplitudes required, e.\ g.\
in coherent control calculations of laser-matter interactions within TDDFT \cite{werschnik}, currently 
a hot topic since atto-second laser pulses allow the control of the electron dynamics. 
This poses one fundamental question: 
How can state-to-state transition probabilities be extracted from TDDFT?\\ 
The ultimate goal of the study of the in general non-linear response of the many-body 
system to a time-dependent perturbation is the determination of 
the state-to-state transition 
amplitude 
\begin{equation}
\label{eq:1}
S_{if} = \lim_{t \to \infty} \langle \chi_f |U (t, -t) | \chi_i \rangle \, ,
\end{equation}
where $|\chi_{i,f} \rangle$ are the initial (final) channel states of the system 
prior to $(i)$ and after $(f)$ the perturbation, and $U(t_1, t_2)$ is the time 
evolution operator of the system. The challenge is, thus, to construct 
a functional $S_{i,f} [n]$ that allows to extract $S_{if}$ from TDDFT. 
The present paper addresses 
methods to extract transition probabilities between discrete states of the 
many-body system. As point of reference, we investigate first the evaluation of 
eq.\ (\ref{eq:1}) employing Kohn-Sham orbitals in close analogy to the time-dependent 
Hartree-Fock (TDHF) method (see \cite{bonche,griffin,alhassid,negele} and references 
therein). This method would be the equivalent of the at least zeroth order of a 
time-dependent many-body perturbation theory ($S$-matrix theory) in which
the time-dependent non-interacting Kohn-Sham system is considered as the unperturbed system. 
This method involves three steps of approximations: The initial state, the final state
and the many-body propagator are approximated by their TDDFT equivalents.
We encounter similar conceptual problems (``cross-channel correlations'') as TDHF does.
We then formulate a novel functional that allows the determination of 
$S_{i,f} [n]$ which is shown to be free of these deficiencies. We test the method with 
the help of an exactly solvable two-electron model.
\\ 
We consider an interacting $N$-electron system of Hamiltonian $H_0$ with stationary
eigenstates $\chi_{i,f}$ 
which is subject to a  perturbation $V(t)$ which is switched on at time $t=0$ 
and switched off at time $t=\tau$. The initial state of the system 
$|\chi_i\rangle$  is assumed to be the ground state and evolves according to the 
time-dependent many-body Schr\"odinger equation (in a.u.)
\begin{equation}\label{eq:2}
i\frac{\partial}{\partial t}|\Psi(t)\rangle= [H_0+V(t)]|\Psi(t)\rangle\;\;,
\;|\Psi(0)\rangle=|\chi_i\rangle\;.
\end{equation}
The state-to-state transition amplitude (or S-matrix) from the initial state 
$|\chi_i\rangle$ to a final state $|\chi_f \rangle$ is defined by the overlap 
of the propagated state $|\Psi(t)\rangle$ with eigenstates $|\chi_f\rangle$ of 
the unperturbed system
\begin{equation}\label{eq:3}
S_{i,f} = \lim_{t \to \infty} \langle \chi_f | \Psi (t) \rangle \; .
\end{equation}
For later reference we note that the time evolution of the projection amplitude for $t > \tau$ is given in terms of the eigenenergies of the asymptotic final states, $\varepsilon_f$, by 
\begin{equation}
\label{eq:4}
\langle \chi_f | \Psi (t) \rangle = \exp (-i  \, \varepsilon_t (t - \tau))
\langle \chi_f |\Psi(\tau) \rangle\;.
\end{equation}
Since the perturbation vanishes for $t > \tau$, the state-to-state transition probability 
is given by $P_{i,f}=|S_{i,f}|^2=|\langle\chi_f|\Psi(\tau)\rangle|^2$.
\\
Within TDDFT, the time-dependent density is represented through the time-dependent 
Kohn-Sham spin-orbitals $\Phi_{\sigma, j}(\vec{r},t)$ as 
\begin{equation}\label{eq:5}
n(\vec{r},t)=\sum_{\sigma=\uparrow,\downarrow} n_\sigma (\vec{r},t) = 
\sum_{\sigma=\uparrow,\downarrow} \sum^{N_\sigma}_{j=1}
|\Phi_{\sigma, j} (\vec{r}, t)|^2\;,
\end{equation}
where $N_\sigma$ denotes the number of electrons of spin $\sigma$. The one-particle 
spin-orbitals $\Phi_{\sigma, j}(\vec{r},t)$ evolve according to the time-dependent 
Kohn-Sham equation
governed by the one-particle Kohn-Sham Hamiltonian
\begin{eqnarray}\label{eq:7}
H^{KS}_\sigma[n_\uparrow,n_\downarrow]&=&-\frac{1}{2}\vec{\nabla}^2+V_{ext}(\vec{r})+V(\vec{r},t)\nonumber\\
&&+V_H[n](\vec{r},t)+V_{xc}[n_\uparrow,n_\downarrow](\vec{r},t)
\end{eqnarray}
which includes the external one-particle potentials, the Hartree potential and
the exchange-correlation potential.
The initial states $|\Phi_{\sigma,j}(0)\rangle=|\Phi_{\sigma,j}\rangle$ are 
the occupied Kohn-Sham orbitals of stationary ground state density functional theory 
(DFT). Although Kohn-Sham orbitals have, a priori, no physical meaning as single-particle
 quantum states, the Slater determinant of Kohn-Sham orbitals, 
$|\Psi^{TDDFT}\rangle:=\hat{A} |\Phi_{\uparrow,1},..,\Phi_{\uparrow,N_\uparrow},
\Phi_{\downarrow,1},...,\Phi_{\downarrow,N_\downarrow}\rangle$, where $\hat{A}$ 
denotes the operator for anti-symmetrization, may be interpreted as zeroth-order 
approximation to the many-body wavefunction in terms of coupling-constant perturbation 
theory \cite{goerlinglevy1,goerlinglevy2}. It is therefore tempting to determine, in 
analogy to the TDHF approximation \cite{bonche,griffin,alhassid,negele},
an approximate S-matrix as the projection amplitude $S_{i,f}(t)\simeq\langle\chi_f|\Psi^{TDDFT}(t)\rangle$.
A delicate question arises at this point: Which are the appropriate channel 
states $\chi_{i,f}$ to project on? For the initial state, stationarity of the propagation
of the system in the limit of a vanishing external perturbation $(V(\vec{r},t)=0)$ mandates 
that $\chi_i$ is a Kohn-Sham Slater determinant of the occupied ground-state orbital. No 
such restriction is imposed on $\chi_f$ when the evolution is calculated by 
forward-propagation. The simplest choice for channel states $|\chi_f\rangle$ are Kohn-Sham 
Slater determinants built up from occupied and virtual Kohn-Sham orbitals 
$|\Phi_{\sigma,j}\rangle$  of the ground state DFT problem, i.e.\ 
\begin{equation}\label{eq:9}
S_{if} \cong  \lim_{t \rightarrow \infty} \langle \chi_f^{DFT} | \Psi^{TDDFT} (t) \rangle \, .
\end{equation}
Reliable transition probabilities can only be expected if both, time-dependent
and stationary Kohn-Sham Slater determinants are good approximations for the
time-dependent and stationary many-body wavefunctions, respectively.
Excited states often show a higher degree of correlation and a single Kohn-Sham Slater 
determinant is no longer a satisfactory approximation. In those cases, more elaborate 
final-state channel functions are needed. Alternatively, 
configuration-interaction (CI) or multi-configuration Hartree-Fock (MCHF) can be employed
\cite{meyer,zanghellini}. Within TDDFT, linear response theory allows the calculation of 
the excitation spectrum by including particle-hole excitations \cite{casida,lr}. As a 
by-product, improved excited-states, i.e.\ the particle-hole reduced density matrix, are 
generated in terms of an expansion in single-particle excitations of Kohn-Sham orbitals. 
In the present case of a two-electron system this approach should give an improved 
wavefunction compared to the initial single Kohn-Sham Slater determinant \cite{rubio}. 
As discussed below, also this approach suffers from the same short-comings as does 
eq.\ (\ref{eq:9}). \\
We therefore introduce a new functional which 
depends only on the time-dependent density. For simplicity and in line with the present 
model system, the derivation is presented for two-electron systems. Generalizations to 
arbitrary many-electron systems are straight forward.
\begin{figure}
\includegraphics[width=12cm]{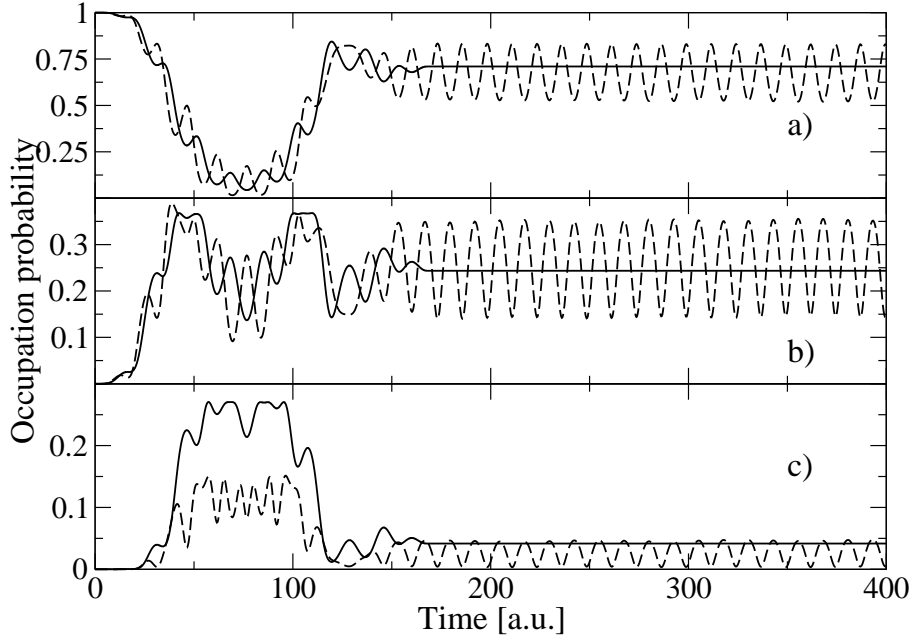}
\caption{Comparison of occupation probabilities of the ground state (figure a),
first excited state $|N_{cm}=1,n_{rel}=0\rangle$ (figure b) and second excited state 
$|N_{cm}=2,n_{rel}=0\rangle$ of exact 
calculation (full line) and TDDFT calculation (dashed line) for a confining frequency 
$\omega=0.25$.
TDDFT occupation probabilities are obtained using the approximate S-matrix 
eq.\ (\ref{eq:9}),
final channel states are Slater determinants of Kohn-Sham orbitals $|0,0\rangle$
,$|0,1\rangle$ and $|0,2\rangle$.
Parameters of the laser pulse: $F_0=0.07$, $\omega_L=0.1839$, $\tau=168$
}\label{fig1}
\end{figure}
Starting point is the expansion of the exact time-dependent wavefunction in terms of a complete set of final-state wavefunctions
\begin{equation}
\label{eq:10}
\Psi (\vec{r}_1, \vec{r}_2, t)  =  \sum_f
\langle \chi_f | \Psi (t) \rangle \chi_f (\vec{r}_1, \vec{r}_2) \, .
\end{equation}  
Using the stationary one-particle reduced density matrix
\begin{equation}
\label{eq:11}
\rho^{(1)}_{f',f}(\vec{r})=
2\int d\vec{r_2} \ \chi_{f'}^*(\vec{r},\vec{r}_2)\chi_f(\vec{r},\vec{r}_2)
\end{equation}
and the time-dependent transition density matrix defined by 
\begin{eqnarray}\label{eq:12}
T_{f',f} (t)= \langle \chi_{f'} | \Psi (t) \rangle^* 
\langle \chi_f | \Psi (t) \rangle
\end{eqnarray}
the exact time-dependent density is given by 
\begin{equation}\label{eq:13}
n(\vec{r},t)= \sum_{f,f'}T_{f',f}(t)\rho_{f',f}^{(1)}(\vec{r})= 
\mbox{Tr}\left[T(t)\rho^{(1)}(\vec{r})\right]\;.
\end{equation}
The transition density matrix and , in particular,
transition probabilities $|S_{i,f}|^2$ can be directly determined by inversion of 
eq.\ (\ref{eq:13}). Our primary interest lies in the diagonal elements $(f'=f)$ of 
the transition density matrix $T_{f',f}(t)\rightarrow S_{i,f'}^*(t)S_{i,f}(t)$ at 
times after the switch-off of the external perturbation $t>\tau$. In this case the 
inversion problem of dimension $N_F \times N_F$ ($N_F$: dimension of truncated 
final-state space considered) can be drastically simplified. Using eq.\ (\ref{eq:4}), 
for non-degenerate final states $(\varepsilon_f \neq \varepsilon_{f'})$, 
the transition probabilities $T_{ff}$ can be extracted from a time-average over an 
interval $(t-\tau)|\varepsilon_{f'}-\varepsilon_f|\gg2\pi$,
\begin{equation}\label{eq:15}
\bar{n}(\vec{r},t):=
\int_{\tau}^{t}\frac{n(\vec{r},t')}{t-\tau}dt'=
\sum_{f,f'}\rho_{f',f}^{(1)}(\vec{r})
\int_{\tau}^{t} \frac{T_{f',f}(t')}{t-\tau}dt'\;,
\end{equation}
leading to the read-out functional
\begin{equation}\label{eq:16}
\lim_{t\rightarrow \infty}\bar{n}(\vec{r},t)=
\sum_{f}\rho_{f,f}^{(1)}(\vec{r})|S_{fi}|^2\;.
\end{equation} 
Unlike eq.\ (\ref{eq:13}), eq.\ (\ref{eq:16}) requires only an $N_F$-dimensional inversion. In practice, the application of eq.\ (\ref{eq:16}) requires evaluating the 
final state densities $\rho_{f,f}^{(1)}(\vec{r})$ at 
$N_F$ distinct points $\vec{r}_j$ $j=1,...,N_f$, so that the matrix 
$R_{f,j}:=\rho_{f,f}^{(1)}(\vec{r_j})$ does not become near-singular and remains 
invertible. The state-to-state transition probabilities 
then become
\begin{equation}\label{eq:17}
| S_{i,f}|^2=\lim_{t\rightarrow \infty}\sum_{j=1}^{N_f} \bar{n}(\vec{r_j},t) 
R^{-1}_{f,j}\;\;\;f=1,..,N_f\;.
\end{equation} 
We have tested the functionals (eqs.\ (\ref{eq:9}) and (\ref{eq:17})) for an exactly 
solvable system of two electrons confined to a harmonic quantum dot 
\cite{taut1,laufer}. In its present 1D version, the electron-electron interaction 
must be replaced by a soft Coulomb potential \cite{grobe}. The Hamiltonian of
the system is given by
\begin{equation}\label{eq:18}
\hat{H}(t)=\sum_{i=1,2}\left(\frac{\hat{p}_i^2}{2}
+\frac{\omega^2}{2}x_i^2-F(t)x_i\right)
+\frac{1}{\sqrt{b+(x_1-x_2)^2}} \, ,
\end{equation}
where $x_i$ and $p_i$ are the coordinates and momenta of electron 
$i$ ($i=1,2$). The softening parameter $b$ is set to $b=0.55$. Similar
systems to describe helium in one dimension were studied in the past 
\cite{grobe,lappas,zanghellini}.
The laser field $F(t)$ with driving frequency $\omega_L=0.1839$ and a peak field 
amplitude $F_0=0.07$ is treated in dipole approximation. A pulse length $\tau = 168$ 
was chosen with a two cycle turn on, a two cycle flat top and a two cycle turn off.
Introducing center of mass (c.o.m.)\ coordinates $R=x_1+x_2$ and 
relative coordinates $r=x_1-x_2$ the Schr\"odinger equation (\ref{eq:2})
can be separated, since the Hamiltonian of eq.\ (\ref{eq:18}) splits into
$\hat{H}=\hat{H}_{rel}+\hat{H}_{cm}(t)$. The eigenstates of the unperturbed system 
are characterized by the set of quantum numbers $(N_{cm},n_{rel})$, the number of 
nodes of the c.o.m.\ and relative wavefunction.
Initial state is the spin-singlet ground state 
$|N_{cm}=0,n_{rel}=0\rangle$. The time dependence of the total Hamiltonian is confined 
to the c.o.m.\ Hamiltonian. The exact time-dependent wavefunction therefore separates 
into a time-dependent  c.o.m.\ and a time-independent relative part, 
$\Psi(r,R,t)=g(r)h(R,t)$. Since the system starts out from the ground state, $h(R,t)$ 
represents a coherent state of the one-dimensional harmonic oscillator driven by an 
external electric field. The dynamics of the density of this model system  
is subject to the harmonic potential theorem \cite{dobson,vignale}, i.e.\ the 
density is rigidly shifted without any distortion.
\\
\begin{figure} 
\includegraphics[width=12cm]{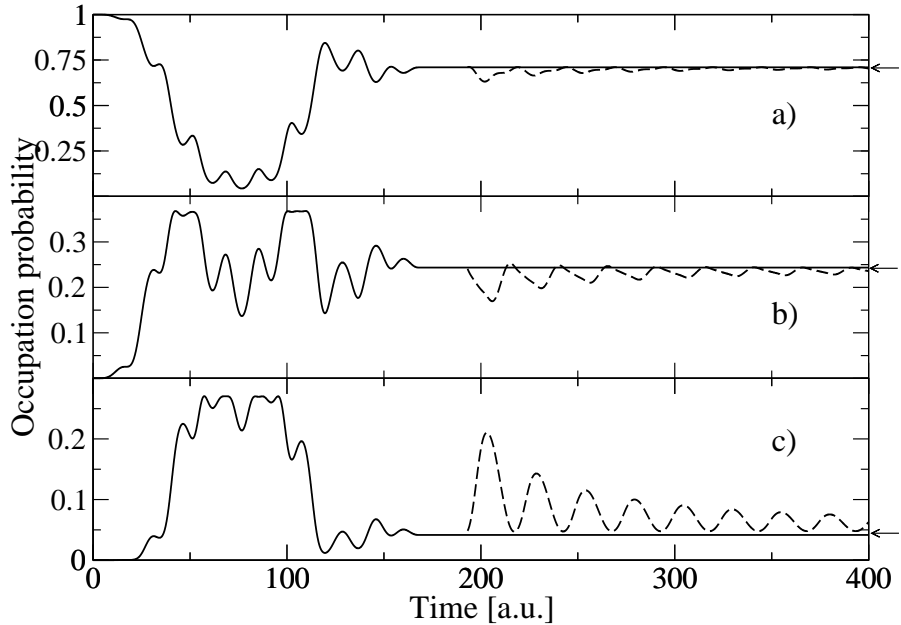}
\caption{Comparison of exact occupation probabilities (solid line) and transition 
probabilities obtained by inversion of eq.\ (\ref{eq:17}) 
(dashed line) for the ground state (figure a), first excited state 
(figure b) and second excited state (figure c).}\label{1dauslese}
\end{figure}
In the case of a two-electron spin-singlet system evolving from the ground state,
the time-dependent Kohn-Sham scheme consists in solving the Kohn-Sham equation
for one doubly-occupied Kohn-Sham orbital $\Phi(x,t)$. For spin-unpolarized two-particle systems
the exact $V_{xc}(t)$ can be constructed using the exact density derived 
from the Schr\"odinger equation and inverting the Kohn-Sham equation \cite{amico,lein}.
Since the harmonic two-electron quantum dot satisfies the harmonic potential theorem
the time-dependence of the exchange-correlation potential is a rigid shift and 
the exact $V_{xc}(t)$ is easily constructed.
We also employed the adiabatic local spin density approximation (ALSDA) with 
self-interaction correction (SIC) \cite{perdew}. Since no reliable correlation 
potential is available for a one-dimensional electron system we only consider
exchange. The $L_1$ norm of the deviation between the exact and TDDFT density is about 
0.2 for the highly correlated system of $\omega=0.25$.
The ground state Kohn-Sham equation generates a set of excited virtual Kohn-Sham 
orbitals $|n\rangle$. Figure 1 shows a comparison of exact occupation probabilities and 
those obtained from the approximate S-matrix (eq.\ (\ref{eq:9})) by projecting 
$\Psi^{TDDFT}(t)=\Phi(x_1,t)\Phi(x_2,t)$ onto Kohn-Sham Slater determinants. 
Shown are the occupation of the ground state (figure a, projection
onto the Kohn-Sham determinant $|0,0\rangle$), the first excited state 
(figure b, projection onto $|0,1\rangle$) and the
second excited state (figure c, projection onto $0,2\rangle$). 
The second excited state $|N_{cm}=2,n_{rel}=0\rangle$ involves a configuration mixture of at least two Kohn-Sham Slater determinants to be well represented (configurations $|0,2\rangle$ and 
$|1,1\rangle$). Neither the projection onto a single Kohn-Sham configuration state 
$|0,2\rangle$ (figure 1c) nor the projection onto the exact excited state (not shown) 
yields satisfactory transition probabilities.
After the switch-off of the laser, the density undergoes 
oscillations resulting in time-dependent Hartree and exchange correlation potentials 
which give rise to oscillations in the occupation probabilities. These are the signatures
of the ''spurious cross-channel correlations'' well-known from TDHF 
\cite{griffin,alhassid}. For the present system, exact excited final states can be easily calculated. 
We have checked that cross-channel correlations persist when we project 
onto exact exit-channel states rather than Kohn-Sham determinants. Moreover, we also 
tested channel states obtained from a TDDFT linear-response equation.
In this equation the exchange-correlation kernel of TDDFT was approximated by
the exchange-only time-dependent optimized effective potential \cite{ullrich,lr} and 
the exact Kohn-Sham orbitals obtained by the exact ground state exchange-correlation 
potential have been used. The obtained excited-state wavefunctions, 
however, do not significantly differ from the single Kohn-Sham Slater determinants 
although the excitation energies are considerably improved compared to the 
Kohn-Sham energy differences. With presently available approximations to the
exchange-correlation kernel, TDDFT linear-response theory does not provide improved
excited-state wavefunctions. We thus conclude that the projection amplitude of 
eq.\ (\ref{eq:9}) is not well-suited to determine an approximate S-matrix within TDDFT.
\\
To test the newly proposed read-out functional which depends only on the density,
we use the exact time-dependent density $n(x,t)$. In this way, errors due to
the approximate exchange-correlation potential can be ruled out and the quality of the 
proposed functional for state-to-state transition probabilities can be 
directly assessed. The sum in eqs.\ (\ref{eq:15}), (\ref{eq:16}) and (\ref{eq:17}) is 
truncated after the 
second excited state. Figure \ref{1dauslese} shows a comparison of transition 
probabilities obtained by projecting the wavefunction according to eq.\ (\ref{eq:3}) 
(solid line) and by the new density functional of eq.\ (\ref{eq:17}) (dashed line) 
for the three lowest lying states. In the limit of $ t \rightarrow \infty$, i.e.\ as the 
averaging interval in eq.\ (\ref{eq:15}) increases, the transition probability converges 
within the numerical accuracy towards the exact result. Numerical errors are due
to the truncation of the sum over final states in eq.\ (\ref{eq:17}). Note that simply averaging over 
the cross-channel correlation 
in eq.\ (\ref{eq:9})(see figure \ref{fig1}) would lead, in general, to incorrect results.
Similar good agreement can be found for other initial states and other systems \cite{nina}.
\\ 
In conclusion, we have investigated two different methods to extract state-to-state transition
 amplitudes from TDDFT calculations. In the first method, the correlated many-body 
wavefunction is approximated by a Slater determinant of the time-dependent
Kohn-Sham orbitals. This approximate wave-function is projected onto appropriate
final states (exact states or Kohn-Sham configuration states). The resulting state-to-state transition probabilities suffer oscillations after the switch-off of the external perturbation. The second read-out functional to calculate state-to-state transition probabilities directly involves the time-dependent densities and represents thus a well-suited density functional within the framework of TDDFT. The problem of cross-channel correlations can be avoided and well-defined transition probabilities can be determined in the asymptotic limit $t\rightarrow \infty$. First results
obtained by evaluating the read-out functional with densities resulting from the 
exact solution of a one-dimensional two-electron Schr\"odinger equation are very 
promising.
The application of the read-out functional to double excitation of helium are currently being
investigated. Work supported by FWF-SFB 016.
\end{document}